# Use of Power Flow Controllers to Enhance Transmission Network Utilisation on the Irish Transmission Network

A. SOROUDI, B. O'CONNELL
J. KELLIHER, J. O'SULLIVAN
EirGrid,
EirGrid, Ireland

B. KELLY, M. WALSH
F. KREIKEBAUM, Y. MEYER

Ireland & USA

## SUMMARY

The Ireland and Northern Ireland power system is in a period of rapid transition from conventional generation to renewable generation and has seen a rapid increase in large demand sites requiring connection into the backbone transmission system. The role of EirGrid and SONI as Transmission System Operator in Ireland and Northern Ireland is to operate, maintain and develop the electricity transmission network. EirGrid ensure that new transmission projects are developed in a way that balances technical, economic, community and other stakeholder considerations. This has resulted in much more detailed evaluation of planning options to maximize utilisation of the existing network which may include but are not limited to uprating the capacity of the existing transmission system mainly through the thermal or voltage uprate of existing circuits. Another method to increase network utilisation is to strategically deploy Power Flow Control devices to relieve system overloads and maximise network transfer capacities. EirGrid has developed a new grid development strategy which places particular emphasis on identifying technologies to help resolve network issues. This paper presents the study findings for application of power flow controller (PFC) to relieve system issues.

Alireza.soroudi@ucd.ie

# 1. INTRODUCTION

The last decade has seen a major increase in the levels of electricity generated from renewable sources like wind. While this is a welcome step away from carbon, this change creates new challenges when operating the electricity grid. Electricity from renewable sources is not always available when and where it's needed. Also, unlike legacy forms of generation, renewables don't deliver a predictable, continuous flow of power. This creates new challenges to managing power flow on grids all over the world.

This evolution in electricity generation comes at a time of other changes in the sector. There is rapid growth in the capacity demanded by high-tech customers like data centres. There is also the potential for rapid change in domestic consumption patterns due to electrification of heat and transport. Accompanying these seismic shifts, there has been a sea change in public acceptance of new electricity infrastructure. The path to acceptance and consent is now long and uncertain - a challenge that is common to all large-scale infrastructure projects. Grid operators need to become more efficient. They also need to add flexibility and responsiveness to existing transmission infrastructure. EirGrid, in particular, made a public commitment about these issues in their 2014 Grid Development Strategy [1]. They committed to develop and improve the grid while minimising the impact on local communities.

Power Flow Control (PFC) is a suitable technology to help increasing network capacity while minimising impact on local communities. PFC is capable of making the transmission network more flexible by changing (increasing/decreasing) the line reactance or by altering the system phase angle. New developments in this technology are allowing utilities to increase network capacity with PFC through direct installation on existing transmission lines or towers and substation deployments are becoming more compact and cost effective. These types of PFC solutions can allow the utility to provide additional reliable network capacity with minimal impact on local communities. Iterations of this type of PFC solution can also be scaled up or down as a need on the network materialises, providing a lower risk solution option to traditional alternatives such as building new overhead lines.

However, a more holistic approach to the optimal location and deployment of PFC technology has not been carried out to date. This paper focuses on an assessment as to where the benefits of PFC can be realised on the transmission network.
The study uses the following high level approach:

1. Assess the level of thermal constraints on each line of the network – thermal loading and duration (number of hours) over a 1 year period;
2. Determine the size and level of PFC required to alleviate thermal overloads ensuring other lines are not subsequently overloaded; and
3. Rank lines where PFC can provide an alternative solution to re-conductor or new line build.

The results of this analysis will allow the TSO to assess PFC solutions more robustly than simply on a case-by-case basis. The analysis also allows the TSO to assess the optimal level of power flow control that is required on each line to remain within Operational and Planning standards over the 1-year study period. In parallel to this analysis, EirGrid has piloted a number of PFC technologies on their network in recent years. The outputs of this analysis will allow the TSO to quickly move into a deployment phase on their network.

## 2. METHODOLOGY

### 2.1 Overview

A new methodology and tool have been developed for assessing the optimal locations for PFC technology on the Irish transmission network. The analysis involves performing Techno-Economic and Technical Load Flow Study analysis in order to identify potential locations to deploy PFC on the network. The high level methodology taken to complete the study is shown out in Figure 1.

The tool performs two key studies to help evaluate deployment options of PFC on the Irish Transmission Network.

- **Techno-Economic Study:** This study creates a year of hourly dispatches as an input into the technical study.
- **Technical Load Flow Studies:** Load flow contingency analysis is performed on each dispatch from the techno-economic model. This is used to determine the locations and amount of PFC required on specific circuits.

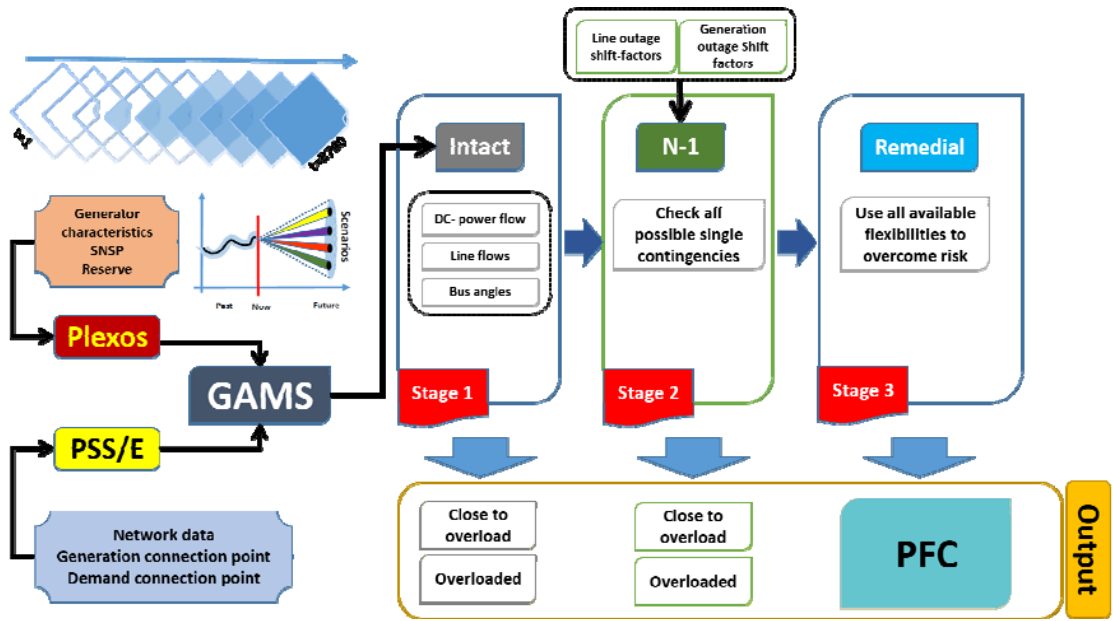

**Figure 1: High Level Methodology for determination of PFC deployment locations on the Irish Transmission Network**

### 2.2 Techno-economic study

The techno-economic analysis uses a production cost modelling approach in PLEXOS. The techno-economic model applies generation and system security constraints to an electricity market model. It does not consider any transmission network elements during the optimisation. The technical and economic characteristics of the generating units (thermal and renewable generators) are used for its techno-economic analysis. This includes the following data:
- Generation Portfolio and Technology Type;
- Generator Short Run Marginal costs;

- RES Availability Curves (including wind and solar);
- Demand Values and Variation pattern;
- Operational Constraints (such as System Non-Synchronous Penetration (SNSP));
- Future Energy Scenarios [2] (taking into demand growth rate, RES penetration level, new generation/demand connections).

The output of this model is an hourly dispatch for 8,760 hours that can be used as an input to the technical load flow studies to identify network needs.

## 2.3 Technical Load Flow Studies

The technical load flow analysis was performed using the TSO developed Power Flow Control Analysis Tool (PFCAT). The General Algebraic Modelling Systems (GAMS) program was chosen as the base software to be used for the analysis [3]. This mathematical tool is used for solving complex linear and non-linear optimization problems. Traditional power flow equations were input into GAMS to carry out DC load flow analysis.

The tool imports the techno-economic hourly generation dispatch for the year, a yearly load profile and the transmission network model. It uses these inputs to generate power-flows for each hour of the year and runs contingency analysis on the same network.
This tool methodically progresses through three main calculation steps or stages as part of the analysis, as per Figure 1:

### 2.3.1   Stage 1 - Intact Network

The power flow analysis is carried out for every single hour of the 1 year period in question. It is assumed that the network is intact. No outage or maintenance schedules inputs are assessed. The DC power flow simulation is performed and all the line flows are calculated and checked against their corresponding line ratings. The following conditions are recorded by the tool:

1. Lines that have power flows over 100% of its seasonal rating - Duration and Severity of overload; and
2. Lines that are close to being in overload condition. That is when line power flows are more than 90% and less than 100% of its seasonal rating.

The tool records these conditions in a workbook for the year of interest. As a DC power flow was used in this analysis, the seasonal ratings of the lines are reduced by 10% to account for the reactive power flows that are not calculated in a standard DC analysis.

### 2.3.2   Stage 2 - Single Contingency Analysis (N-1)

The Transmission System Security Planning Standards [4] dictate how the transmission system is designed. The system shall be designed to withstand the more probable contingencies without widespread system failure for loss of a single piece of plant. A single contingency analysis (N-1) is performed on the network using the tool. All of the line flows are checked against the line rating as in Stage 1, The tool also calculates generation shift factors and line outage shift factors, which are critical to determining the location for PFC deployment.

- **Generation Shift Factor (GSF)** indicates the influence of generation output change on the flow of each line. These factors are dependent on the network topology and do not change by operating point of generation or demand profile;
- **Line Outage Shift Factor (LOSF)** indicates that if a line is out of service (due to a contingency), what would be the power flow impact of this outage on other existing lines. These factors are dependent on the network topology and do not change by operating point (generation and demand profile).

The calculation of GSF and LOSF allows the algorithm to use the results obtained in Stage 1 and avoid heavy computational burden of running N-1 analysis for 8,760 periods. The Stage 2 load flow studies identifies any lines that are overloaded or at risk of overload in case of N-1 contingencies. Since the GSF and LOSF are not dependent on the operation point of the system, this assumption is valid in the context of identifying the location for PFC.
The lines that are identified using the results obtained at Stage 1, will be checked in Stage 2 and 3. Additionally, the severity of the overload state and the number of elements that can cause overload are checked and recorded.

### 2.3.3 Stage 3 – Power Flow Control Assessment

In Stage 3, remedial actions using power flow controllers to resolve the overload condition on identified lines of interest from Stage 1 and Stage 2 are identified. In order to relieve the overload on these lines, the line reactance is increased, and the improvement against the line thermal rating is calculated. The line reactance is only increased by up to 40%; therefore in some instances, the line overload may not be fully mitigated by increasing the reactance alone. After the addition of PFC on a line, the tool performs checks on every line to ensure that there is no negative thermal impact on other lines. Other remedial actions can be developed in the tool such as rescheduling the thermal and renewable generating units, but this was not addressed in this analysis.

Performing the analysis in this way allows the TSO to assess the severity of potential overloads on the network and benefits of using PFC to alleviate those overloads for the year in question. Further improvements to include Outage and Maintenance schedules, which will increase the need further for PFC are being considered by the TSO. The algorithm, was configured to assess lines at 110kV voltage level only. Further analysis will look at the potential to deploy PFC at higher voltage levels.

## 3. SIMULATION RESULTS

### 3.1 Stage 1 and Stage 2 Results

A thorough assessment was carried out on the benefits of PFC on Irish transmission network over a one year period. The algorithm identified a number of lines on the Irish Transmission Network which were overloaded or close to overload under N-1 contingency and these are depicted in Figure 2. These are categorised by region in the diagram also. No overloaded conditions were identified in Stage 1 (intact network), meaning al critical line overloads were identified as part of the Stage 2 analysis. Each identified line is distinct in terms of the degree of overload, the number of overloaded hours, and number of contingencies that might cause an overloaded situation.

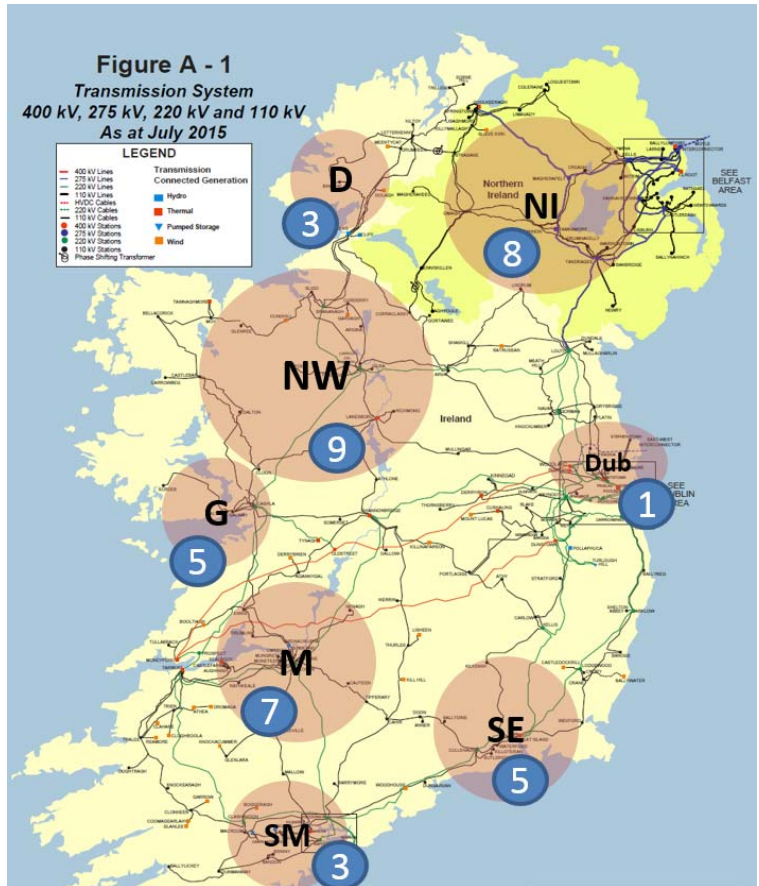

**Figure 2: Map showing number of overloaded lines by region for the 1 year study period**

It can be seen that the number of overloaded lines is highest in the West of Ireland as this is the location of a number of new renewable energy generation customers. This is to be expected given the growth of generation in this area.

Outputs of the tool include the duration of overloads on all circuits and also the severity of overloads. The number of overloaded hours for each line is shown in Figure 3.

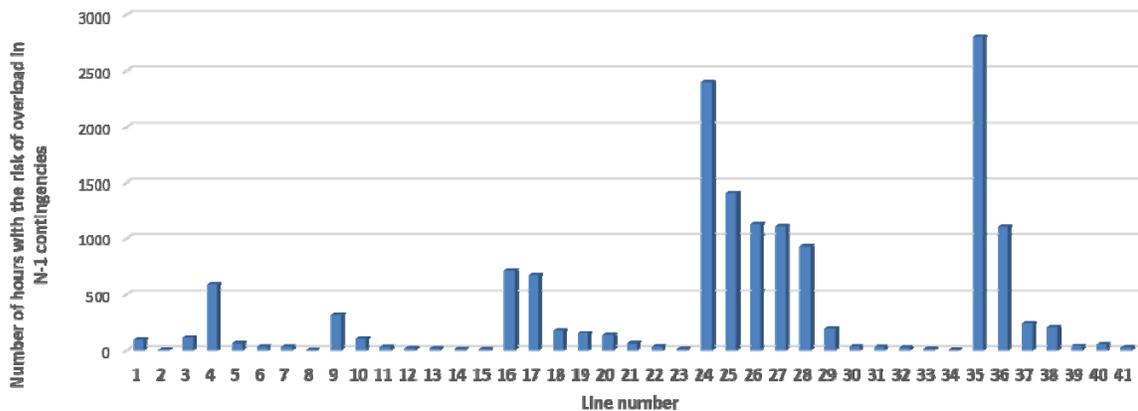

**Figure 3: Duration for which lines were overloaded as identified in Stage 2**

Alireza.soroudi@ucd.ie

## 3.2 Stage 3 Results

The lines identified as being overloaded or close to overloads in Stage 1 and Stage 2 are assessed to determine whether a PFC solution would relieve the overloads. The results of this analysis can be categorised into three groups as follows:

### 3.2.1 Overload Fully Resolved

The overload situation of lines in this category can be fully resolved by increasing the line reactance while also not causing any other lines to get overloaded. The reason is that the PFC tries to push the power to the parallel paths to reduce the line loading if a parallel path exists. One such example is shown below in Figure 4. It can be observed that for an overload on the Bus 1-Bus 3 110 kV line, two parallel paths exist between Bus 1-Bus 2 and Bus 1-Bus 3 where a PFC solution could be utilised. The tool determines the level of PFC required to resolve the overload while ensuring no negative impact on other lines. The LOSF indicates the line best suited to deploy PFC to alleviate this type of thermal overload issue.

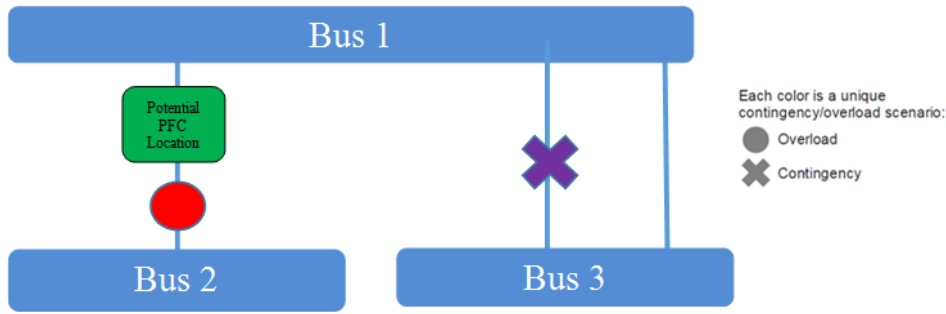

**Figure 4: Sample case where PFC can be used to fully resolve a line overload in case of N-1 contingency**

### 3.2.2 Partially Resolved Overload

The overload situation of lines in this category can be partially resolved by increasing the line reactance. The reason for this is that by pushing the power onto the parallel paths, some other lines, not already overloaded, may become overloaded. It may also deteriorate the overloaded condition of other lines already overloaded in the same region. In Figure 5 it can be seen that Bus 1-Bus 2 is overloaded after a N-1 contingency of Bus 1-Bus 3 line. The line between Bus 3 and Bus 4 was identified as being close to overload condition in the Stage 2 analysis. Deploying PFC on the line between Bus1-Bus 2 causes the line between Bus1-Bus 3 to become overloaded. It is therefore not possible to completely resolve both overloads through a single deployment of PFC. Optimisation of PFC deployment on Bus 1-Bus 2 line and Bus 3-Bus 4 line could potentially resolve the issue. The TSO is expected to investigate multiple deployments of PFC in future analysis to address these kinds of network needs.

Alireza.soroudi@ucd.ie

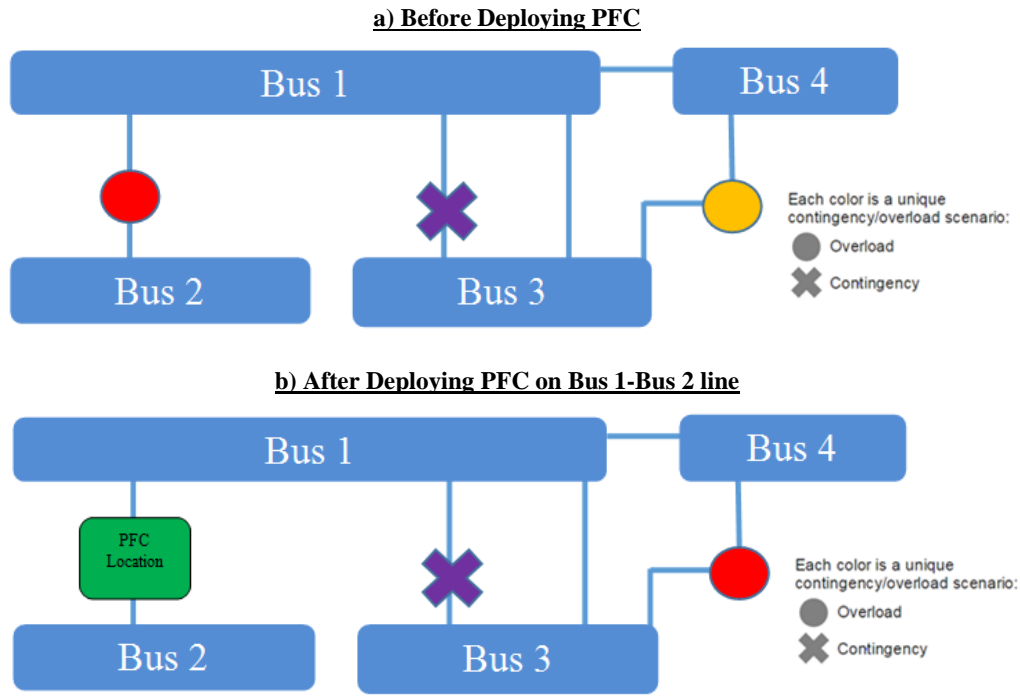

**Figure 5: PFC overloads in case of N-1 contingency**

Another example of Partially Resolved Overload in this context would occur where the deployment of 40% change in line compensation using PFC does not fully alleviate the overload.

### 3.2.3 No Change in Power Flow

It should be noted that the PFC used in this analysis works on the concept of shifting the power to the other parallel lines (i.e increasing line reactance only). If no parallel path exists then the overload or risk of overload cannot be mitigated. There are areas on the Irish transmission network with such characteristics. This is shown in Figure 6, where an N-1 contingency of the line between Bus 1 and Bus 3 overloads the line between Bus 1 and Bus 2. Deploying PFC in this case will not resolve the overload condition.

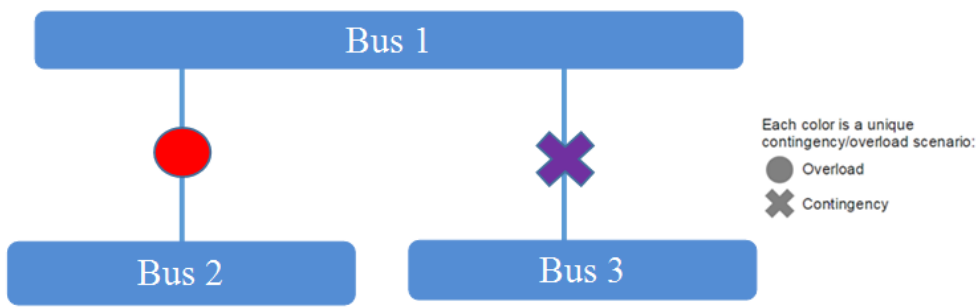

**Figure 6: Condition where PFC cannot resolve the possible overload under N-1 contingency scenario**

Alireza.soroudi@ucd.ie

### 3.3 Performance of PFC at Alleviating Overloads

Applying a 40% change in reactance on the lines of interest was highly effective in fully resolving overloads in a number of cases on the Irish Transmission Network. The percentage of line overloads that were fully or partially resolved through the deployment of PFC is shown in Figure 7b.

It can be seen that in some cases, lines which were overloaded for high number of hours throughout the study year period were fully resolved through the deployment of PFC. For example, Line 35 showed overload condition for 2750 hours over the year and in each period, deploying PFC could alleviate that overload for approximately 50% of the time. That line would be considered a candidate to deploy PFC in the short term.

a) **Severity of overloads on Irish Transmission network**

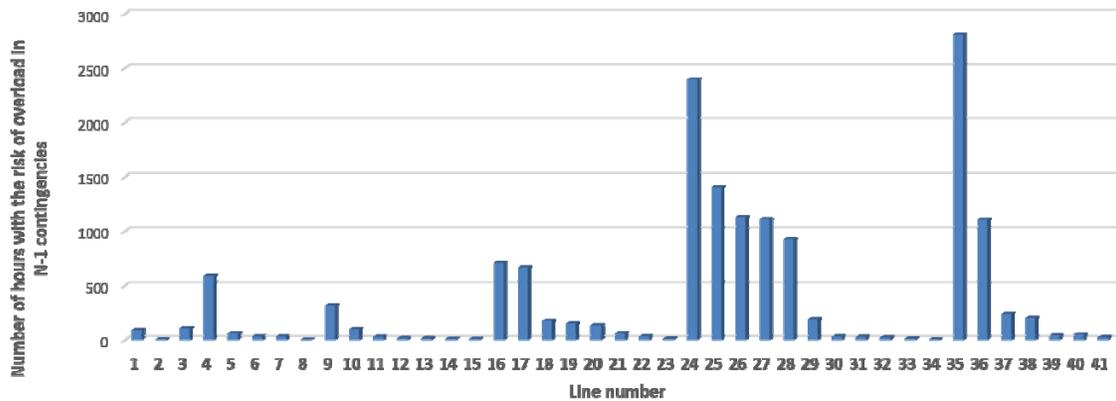

b) **Impact of PFC in mitigating line overloads on Irish Transmission Network**

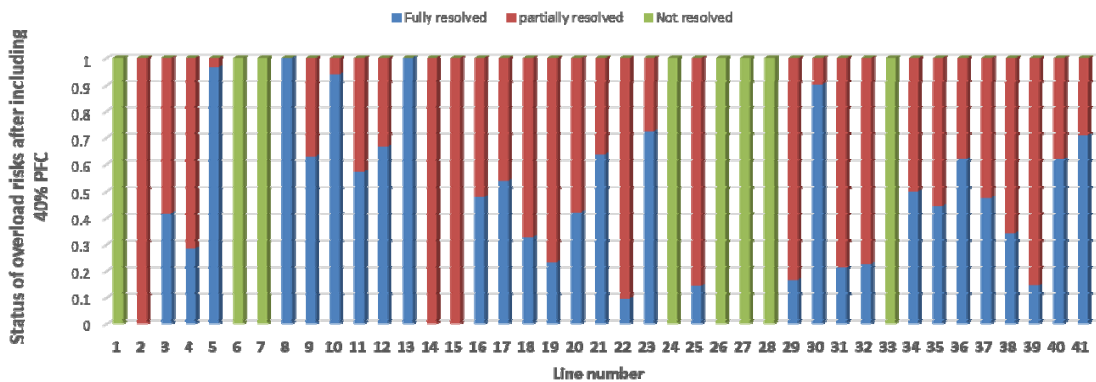

**Figure 7: Performance of PFC in alleviating overloads on Irish Transmission Network**

## 4. CONCLUSION

An innovative planning tool has been developed to investigate the role of Power Flow Controller's in enhancing the utilisation of existing transmission network. The developed tool identifies the deployment location and scheduling of PFC to maintain the safe and efficient operation of existing transmission network. The study identified a number of potential lines to deploy PFC on to help alleviate thermal overloads and also categorised the risk that these lines would be overloaded based on the expected duration of the overload.

Plans to carry out further analysis to build on the results presented in this paper include:

1. Increasing the level of line compensation in future studies – This analysis assumed that PFC would provide a maximum impedance change of 40%. Further analysis will investigate higher levels of power flow control, which will substantially increase need for PFC on the network.
2. Assessing line overloads on the EHV Network (220 kV and 400 kV) – The candidate lines assessed in this paper were 110 kV overhead lines only. However the developed methodology can be easily scaled to assess issues at higher voltage levels.
3. Include a Techno Economic Assessment of PFC over standard redispatch schemes to determine an optimal investment decision – The focus of this paper was on the technical aspects of PFC to resolve thermal overloads and economic assessments of alternative solutions was not carried out.
4. Carry out the study over a longer planning horizon – The 8,760 hour study was carried out for one year only, 2020.
5. Include Outage and Maintenance Schedules – Outage schedules were not included in the analysis.
6. Assessing multiple deployments of PFC to alleviate multiple line overloads was not considered in this analysis.

The analysis has paved the way for a more detailed evaluation of PFC in future and also for identifying use cases to increase network utilisation. The network utilisation is improved since the technical risks can be resolved without building new overhead lines if PFCs are allocated and operated.

**BIBLIOGRAPHY**

[1] EirGrid-Draft-Grid-Development-Strategy, 2014, http://www.eirgridgroup.com/site-files/library/EirGrid/EirGrid-Draft-Grid-Development-Strategy.pdf
[2] EirGrid Tomorow's Energy Scenarios 2017, http://www.eirgridgroup.com/site-files/library/EirGrid/EirGrid-Tomorrows-Energy-Scenarios-Report-2017.pdf
[3] A. Soroudi, "Power System Optimization Modeling in GAMS." (2017).
[4] EirGrid's Transmission System Security and Planning standards, 2016 http://www.eirgridgroup.com/site-files/library/EirGrid/EirGrid-Transmission-System-Security-and-Planning-Standards-TSSPS-Final-May-2016-APPROVED.pdf

Alireza.soroudi@ucd.ie